\documentclass{aastex}          %% The manuscript based on AASTeX v5.x
\usepackage{spr-astr-addons}    %% mimicing ASTR journal style
\begin{document}
%% Article title
%
  \title{The physical fundamental plane of black hole activity: revisited}
%% Running heads

\shorttitle{The physical fundamental plane of black hole activity}

\shortauthors{Liu et al.}

\author{Xiang Liu\altaffilmark{1,2}}
\and
\author{Zhenhua Han\altaffilmark{1,3}}
\and
\author{Zhen Zhang\altaffilmark{1}}

\email{liux@xao.ac.cn}

\altaffiltext{1}{Xinjiang Astronomical Observatory, Chinese
Academy of Sciences, 150 Science 1-Street, Urumqi 830011, PR
China}

\altaffiltext{2}{Key Laboratory of Radio Astronomy, Chinese
Academy of Sciences, Urumqi 830011, PR China}

\altaffiltext{3}{Graduate University of Chinese Academy of
Sciences, Beijing 100049, PR China}

\begin{abstract}

The correlation between the jet power and accretion disk luminosity is investigated for active galactic nuclei (AGNs) and black hole X-ray binaries (BHXBs) from the literature. The power-law correlation index is steep ($\mu \sim$ 1.0--1.4) for radio loud quasars and the `outliers' track of BHXBs, and it is flatter ($\mu \sim$ 0.3--0.6) for radio loud galaxies and the standard track of BHXBs. The steep-index groups are mostly at higher accretion rates (peaked at Eddington ratio $>$ 0.01) and the flatter-index groups are at relatively low accretion rates (peaked at Eddington ratio $<$ 0.01), implying that the former groups could be dominated by the inner disk accretion of black hole, while the jet in latter groups would be a hybrid production of the accretion and black hole spin. We could still have a fundamental plane of black hole activity for the BHXBs and AGNs with diverse (maybe two kinds of) correlation indices. It is noted that the fundamental plane of black hole activity should be referred to the correlation between the jet power and disk luminosity or equivalently to the correlation between jet power, Eddington ratio and black hole mass, rather than the jet power, disk luminosity and black hole mass.

\end{abstract}


\section{Introduction}

Radio jets in AGNs and the black hole X-ray binaries could be produced by the mass accretion onto black hole (e.g.
Blandford \& Payne 1982) and/or by the black hole spin (e.g. Blandford \& Znajek 1977). It is suggested there is a linear positive correlation between radio jet power and disk luminosity in the radio loud quasars which are dominated by the disk accretion (van Velzen \& Falcke 2013). However, for radio galaxies, radio-loud Seyferts and the low luminosity AGNs (LLAGNs) their jet power could be a hybrid production of the disk accretion and black hole spin, because their power-law correlation indices are flatter than the linear correlation expected in the accretion dominated jet (Liu \& Han 2014). In this letter, we investigate the implications of these results to the so-called fundamental plane of black hole activity (see, e.g., Merloni et al. 2003; Falcke et al. 2004; Li et al. 2008).

\section[]{The physical fundamental plane of black hole activity}

Only mass and spin are concerned in astronomical black holes, and the bi-polar jets are formed through accretion disk and/or spin of black hole as mentioned. The disk luminosity defined as $L_{disk}=\varepsilon \dot{M}c^{2}$, with $\varepsilon$ the disk radiative efficiency, being often assumed as a constant for individual working samples, but $\varepsilon$ can be different for different type of sources at significantly different accretion states and which can be much lower than the fiducial 10\% efficiency of a standard thin accretion disk, especially at low accretion rates (Yuan \& Narayan 2014). The jet power defined as $L_{j}=\eta \dot{M}c^{2}$, with $\eta$ the jet efficiency which should be a complicated function of black hole spin, black hole mass, accretion rate and jet orientation angle. The jet formation could either be dominated by the disk beyond a few gravitational radii or dominated by the black hole spin; the former has been investigated by Liu \& Han (2014), in which the jet power can be linearly correlated with the disk luminosity if the black hole spin contribution to jet power is small/neglectable, which is expressed as:

\begin{equation}
L_{j}=(\eta /\varepsilon)L_{disk}.
\end{equation}

It can be rewritten as:

\begin{equation}
L_{j}=1.26\times10^{38} (\eta /\varepsilon) [(L_{disk}/L_{Edd})
M/M_{\odot}] (erg/s).
\end{equation}

This relation suggests a linear proportionality of jet power to the
product of the Eddington ratio ($\lambda=L_{disk}/L_{Edd}$) and
the black hole mass ($M$), with the coefficient of
$\eta$/$\varepsilon$ assumed to be constant for similar type of sources which being at narrow range of accretion rates and the black hole spin contribution to the jet power is small, as the first approximation. The $\eta$/$\varepsilon$ will be otherwise not constancy for different type of sources which being at widely distributed accretion states or the black hole spin contribution is significant to the jet power. It is suggested that the spin-produced-jet power is approximately (Yuan \& Narayan 2014):

\begin{equation}
P_{spin}\sim 2.5 (a)^{2} (\Phi/\Phi_{MAD})^{2} \dot{M}c^{2},
\end{equation}

$a$ is the spin parameter, $\Phi$ is magnetic flux and the $\Phi_{MAD}$ is the upper limit of $\Phi$ for the magnetically arrested inner disk (Tchekhovskoy et al. 2011), i.e. the $\Phi$ would depend on but less than the $\Phi_{MAD}$ (Yuan \& Narayan 2014), for example if we assume a function $\Phi \propto (\Phi_{MAD})^{\kappa}$ with $\kappa<1$, then from the formula of $\Phi_{MAD}$ in Yuan \& Narayan (2014) and the equation 3, we have

\begin{equation}
P_{spin}\propto a^{2}m^{3\kappa-2} \dot{m}^{\kappa}/\varepsilon,
\end{equation}

where $m$ is black hole mass in solar mass unit, $\dot{m}$ is mass accretion rate in Eddington accretion rate unit. We see that in this case, the spin-induced-jet power has a flatter index ($\kappa$) than the linear correlation which is accretion dominated; only in extreme case, i.e. the $\Phi=\Phi_{MAD}$, the spin-induced-jet power has a linear correlation with accretion rate.

We refer to the relation 1 and 2 as the `physical fundamental plane of black hole activity' at the first approximation, which resulted from the physics discussed by Liu \& Han (2014), and it has two parameters: $L_{j}$, $L_{disk}$, or three parameters: $L_{j}$, $\lambda$ and black hole mass $M$. Merloni et al. (2003) first proposed a fundamental plane of black hole activity with the three parameters: $L_{j}$, $L_{disk}$, and $M$, and the correlation between the three quantities was analyzed, but which was not derived from known physics.

\section[]{Statistical properties for AGNs and black hole X-ray binaries}

The fundamental plane of black hole activity was investigated for stellar mass black hole X-ray binaries (BHXBs) and massive black holes in radio loud AGNs, to building up a scaling relation between them, and basically there is a positive power law correlation between jet power and disk luminosity, for both BHXBs and AGNs (see, e.g., Fig.3 of Merloni et al. 2003). There is a gap of jet power between the AGNs and BHXBs, the gap could be filled with intermediate-massive black hole accretions. According to Liu \& Han (2014), the physical relation is the equation 1 and 2, rather than a relation between $L_{j}$, $L_{disk}$, and $M$, and it is not reasonable to add a third parameter (black hole mass $M$) in the equation 1 to unify the BHXBs and AGNs with a single slope as shown in Fig.5 of Merloni et al. (2003). As analyzed from a large sample of AGNs by Li et al. (2008), jet power turns out to be not correlated with black hole mass, but strongly correlated with disk luminosity with the average power-law index of 1.17$\pm$0.42 for high redshift AGNs (from Table 4 of Li et al. 2008). More accurately, van Velzen \& Falcke (2013) demonstrated that there is a tight linear correlation between jet power and disk luminosity for radio loud FRII quasars, which is expected from the equation 1.

However, for radio galaxies, low luminosity AGN and Seyfert galaxies as shown in Liu \& Han (2014), the correlation indices (0.3--0.6) are much flatter than the linear correlation found in the radio loud quasars (e.g., Willott et al. 1999; van Velzen \& Falcke 2013), suggesting a third parameter could work on the jet of these galaxies, which might be the black hole spin as discussed by Liu, Zhang \& Han (2014). As analyzed in previous section, there could be a flatter power-law index between the spin induced jet power and the disk accretion rate. For instance, for a value of $\kappa$=0.5 in equation 4, we have $P_{spin}\propto a^{2}m^{-0.5} \dot{m}^{0.5}/\varepsilon$, and rewritten as $P_{spin}\propto a^{2}m^{-1} L_{disk}^{0.5}/\varepsilon$, this suggests that there is a possibility to explain the flatter slopes (0.3--0.6) in the radio galaxies, Seyferts and LLAGNs with the black hole spin contribution.

Furthermore, it is found that there is a low correlation index (0.43$\pm$0.37) on average for radio quiet AGNs (Li et al. 2008), implying that the black hole spin may dominate the weak radio emission in the radio quiet AGNs.

Actually, for some black hole X-ray binaries, a correlation between jet power and black hole spin has been found (Steiner et al. 2013), this is the first evidence that the jet power is correlated with the black hole spin in BHXBs. The radio jet luminosity also tightly correlates with the X-ray luminosity (which supposed to be the disk luminosity) at the hard/low state of BHXBs, with a correlation index around of $\sim$ 0.5--0.7 (Fender et al. 2004; Gallo et al. 2003). Recent years, different from the standard index of $\sim$ 0.6, some `outliers' with steeper indices of 1--1.4 have been found in BHXBs, in which the radio luminosity is lower but increases more steeply with disk luminosity in the `outliers' than in the standard track, for similar disk luminosity (see, Gallo et al. 2012; Corbel et al. 2013). It is not clear what causes the steepened correlation index in the `outliers', however, with compared to the steep index $\geq$1 in radio loud quasars (Fernandes et al. 2011; Kalfountzou et al. 2012; van Velzen \& Falcke 2013), the `outliers' track in the BHXBs might be related to an accretion dominated jet similar to that in the radio loud quasars.

\section[]{Non-single slope from the black hole X-ray binaries to AGNs}

\begin{figure}
    \includegraphics[width=8cm]{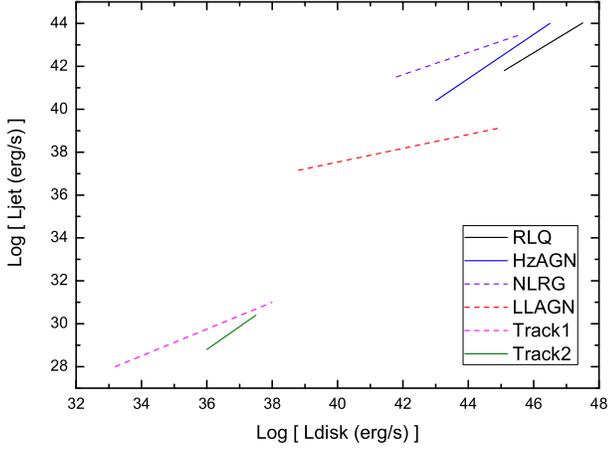}
    \caption{Log [jet power] vs. log [disk luminosity], the best-fitting slope $\mu$=1.0 for the radio loud quasars (RLQ: van Velzen \& Falcke 2013), $\mu$=1.2 for the high-z radio loud AGN (HzAGN: Li et al. 2008), $\mu$=0.52 for the narrow line radio galaxies (NLRG: Liu \& Han 2014), $\mu$=0.32 for the LLAGNs and Seyferts (LLAGN: Liu \& Han 2014), and $\mu$=0.6 and 1.4 for the standard track (Track1) and the `outliers' (Track2) of black hole X-ray binaries (Gallo et al. 2012), have been plotted respectively. The dash lines have the flatter power-law indices in range of 0.32-0.6, and the solid lines show the steep indices of $\mu \sim$1.0-1.4.  }
     \label{fig1}
  \end{figure}

\begin{figure}
    \includegraphics[width=8cm]{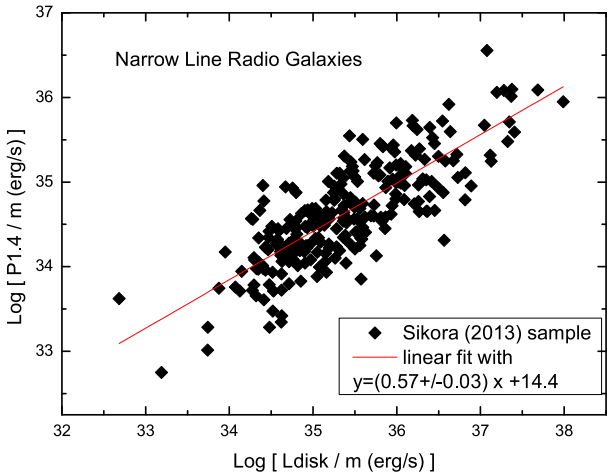}
    \caption{Log [1.4 GHz jet power divided by black hole mass (in solar mass unit)] vs. log [disk luminosity divided by black hole mass (in solar mass unit)] for the narrow line radio galaxies (Sikora et al. 2013; Liu \& Han 2014), the best linear fit is shown.}
     \label{fig2}
  \end{figure}

\begin{figure}
    \includegraphics[width=8cm]{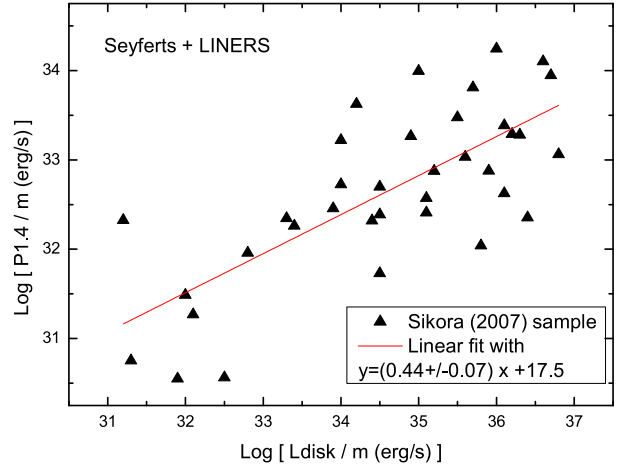}
    \caption{Log [1.4 GHz jet power divided by black hole mass (in solar mass unit)] vs. log [disk luminosity divided by black hole mass (in solar mass unit)] for the Seyferts and LLAGNs (Sikora et al. 2007; Liu \& Han 2014), the best linear fit is shown.}
     \label{fig3}
  \end{figure}

\begin{figure}
    \includegraphics[width=8cm]{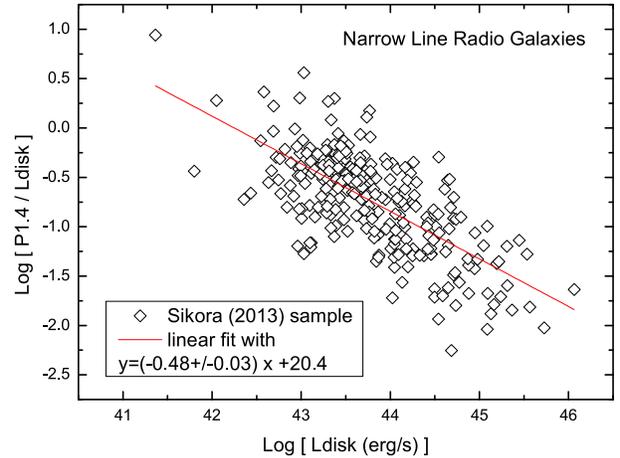}
    \caption{Log [1.4 GHz jet power divided by disk luminosity] vs. log [disk luminosity] for the narrow line radio galaxies (Sikora et al. 2013; Liu \& Han 2014), the best linear fit is shown.}
     \label{fig4}
  \end{figure}

\begin{figure}
    \includegraphics[width=8cm]{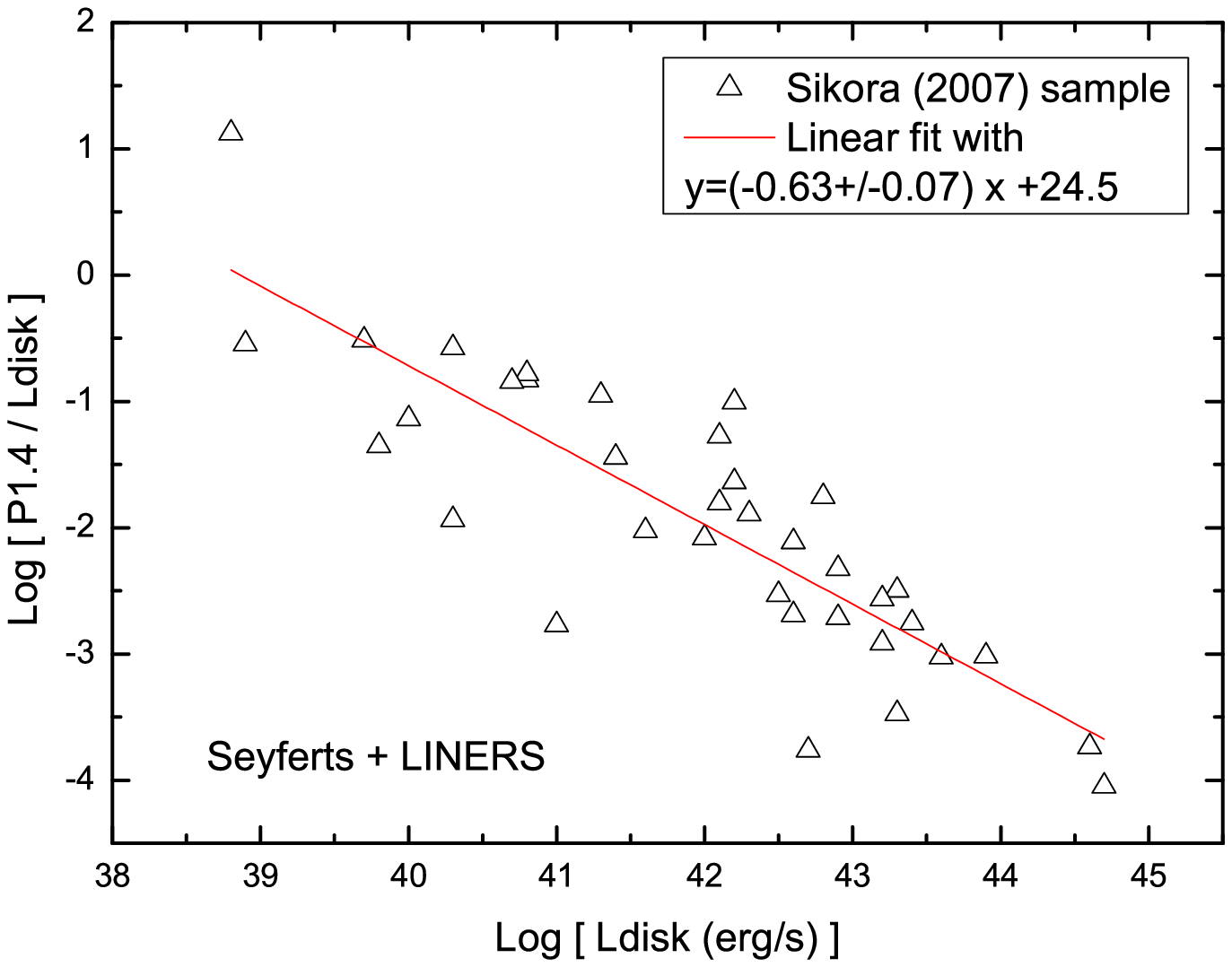}
    \caption{Log [1.4 GHz jet power divided by disk luminosity] vs. log [disk luminosity] for the Seyferts and LLAGNs (Sikora et al. 2007; Liu \& Han 2014), the best linear fit is shown.}
     \label{fig5}
  \end{figure}

As analyzed above, there are different correlation slopes between $L_{j}$ and $L_{disk}$ for different samples of AGNs and `different tracks' of BHXBs. We illustrate the different correlation indices between jet power and disk luminosity in Fig.~\ref{fig1}, in which the slopes for the FRII quasars from van Velzen \& Falcke (2013), high redshift (z$>$0.5) AGNs from Li et al. (2008), the narrow line radio galaxies and LLAGNs/Seyfert galaxies from Liu \& Han (2014), and the black hole X-ray binaries from Gallo et al. (2012) have been plotted respectively, in which the standard track and `outliers' track of BHXBs are marked as `track1' and `track2'. We also tried to fit radio galaxies and Seyferts/LLAGNs with both jet power and disk luminosity divided by black hole mass, the results as shown in Fig.~\ref{fig2} and Fig.~\ref{fig3} indicate that the correlation significance is improved but correlation index is not significantly different from that not divided by the black hole mass in Liu \& Han (2014), implying that the black hole mass plays a minor role in the correlation between jet power and disk luminosity for individual samples. We also plot the jet-disk luminosity ratio versus disk luminosity for the radio galaxies and Seyferts/LLAGNs as shown in Fig.~\ref{fig4} and Fig.~\ref{fig5}, they show similar anti-correlation with compared to the radio loudness versus disk luminosity as analyzed in Liu \& Han (2014), this is due to the correlation index between jet power and disk luminosity is flatter than a linear correlation.

Obviously, in Fig.~\ref{fig1}, it is not possible for us to unify the different slopes with a single power-law slope, i.e. there is no `unified fundamental plane' with a single correlation index from
BHXBs to AGNs. However, there appear two `kinds' of correlation slopes for both the BHXBs and AGNs, one has $\mu \geq$1 ($L_{j}\propto L_{disk}^{\mu}$), the other has $\mu$ = 0.3--0.6. In this sense, we still have a fundamental plane of black hole activity for BHXBs and AGNs with the two kinds of correlation indices: $\mu \geq$1 (the solid lines in Fig.~\ref{fig1}) which could be dominated by the disk accretion, and $\mu \sim$ 0.3--0.6 (the dash lines in Fig.~\ref{fig1}) which would be determined by both the mass accretion and black hole spin. As discussed in Liu \& Han (2014), a minimum accretion rate is required for the black hole spin to create a jet, and a high accretion rate would be able to suppress the spin-induced jet and instead to form a more powerful jet by the accretion itself.

The physical fundamental plane shown in Fig.1, can also be equivalently plotted with the $L_{j}$, $\lambda$, and black hole mass $M$ from the equation 2, by replacing the $L_{disk}$ with the $\lambda M$ in Fig.~\ref{fig1}, it will be the same shape (not shown here) as the Fig.~\ref{fig1}. For the AGN samples, statistically, the jet power has shown no or only weak correlation with black hole mass, e.g. in the radio loud AGNs from SDSS (Li et al. 2008), in the radio galaxies (Sikora et al. 2013; Liu \& Han 2014), and in the LLAGNs and Seyfert galaxies (Ho 2002). This is also true for the BHXBs in Merloni et al. (2003), there is no correlation between jet luminosity and black hole mass. Therefore, one would expect similar correlation indices to those in Fig.~\ref{fig1}, between jet power and disk accretion rate (Eddington ratio), for example see the Fig.3 (right panel) of Merloni et al. (2003). However, it is not reasonable to correlate the jet power with the disk luminosity and black hole mass as done by Merloni et al. (2003), actually, the reasonable correlation is that between jet power, accretion rate and black hole mass which has been discussed in Heinz \& Sunyaev (2003), and also see the equation 2.

\section[]{Discussion and summary}

The Fig.~\ref{fig1} shows diverse correlation slopes between jet power and disk luminosity for AGNs and black hole X-ray binaries, and both the AGNs and BHXBs have two subgroups showing either steep or flatter correlation indices. The AGNs with steep correlation indices have on average higher accretion rates than that with flatter indices. For example, the large sample of AGNs at redshift $>$0.5 analyzed by Li et al. (2008) shows a steep correlation index with Eddington ratio peaking around 0.01-0.1; the van Velzen \& Falcke (2013) sample of radio loud FRII quasars shows a linear correlation with a median Eddington ratio of 0.08 at the redshift around 1.2. For the `outliers' of BHXBs appeared in the high disk luminosity regime of $L_{d} \sim 10^{36-39}$ erg/s, see Corbel et al. (2013), they are at higher accretion rates with the Eddington ratio mostly $\geq$ 0.01 assuming a similar black hole mass of $\sim$ 10 solar masses in the BHXBs. Whereas, the samples with flatter correlation indices between jet power and disk luminosity are mostly at lower accretion rates. For example, the radio galaxies and LLAGNs/Seyferts at redshift $<$ 0.4, are peaked at the Eddington ratio of $<$ 0.01 (Liu \& Han 2014; Sikora et al. 2007, 2013); the standard track of BHXBs with the slope $\sim$ 0.6 has been mostly at lower disk luminosity with the Eddington ratio of $<$ 0.01.

Gallo et al. (2014) find that 24 BHXBs at quiescence state (at an extremely low Eddington ratio) have a radio-X-ray power law index of 0.61$\pm$0.03, which is consistent with the standard track and so probably dominated by the black hole spin. The very weak jets are not relativistic and the particle acceleration is much weaker after the transition into quiescence (Plotkin et al. 2015). Coriat et al. (2011) proposed that the outliers would be consistent with the presence
of radiatively efficient accretion flow (hot accretion flows
or accretion disk-corona), and the standard track would be still in a radiatively inefficient accretion flow (such as ADAF models), see Dong et al. (2014), Meyer-Hofmeister \& Meyer (2014), and Qiao \& Liu (2015) for more comments on this possibility. The difference between the two tracks of BHXBs may be also related to a variation in the viscosity parameter (Xie \& Yuan 2012).

In addition, it is found that the radio-loud narrow-line Seyfert 1 galaxies (NLS1s) are mostly at high accretion rates (e.g. Komossa et al. 2006). Recently, Foschini et al. (2015) has complied a large radio-loud NLS1 sample, and there is a steep power law relationship
between radio jet power and disk luminosity after removing some outlier data points by us. This result implies that the NLS1s are quite different with other low luminosity AGNs and Seyfert galaxies. The jet of NLS1s is likely pointing close to our line of sight and so the properties of NLS1s are quite similar to the blazars -- highly variable AGNs (Yuan et al. 2008). As noted in Liu \& Han (2014), however, the blazars are not good targets for the jet-disk coupling analysis because of their highly Doppler boosted flux, therefore, the correction of the Doppler effect to NLS1s should be made in future. For the lobe-dominated radio galaxies and FRII quasars the jet viewing angles are usually large, so the beaming effect could be small, but this effect should be analyzed further for radio loud Seyferts and LLAGNs in future.

Ghisellini et al. (2014) have found a tight linear correlation between radiative jet power and disk luminosity in Fermi gamma ray blazars, in which the observed radiative jet power has been corrected for the Doppler beaming effect. They think that these high luminosity jet powers in the gamma ray blazars could be produced by black hole spin. We think it is also possible that the linear correlation between radiative jet power and disk luminosity is caused by mass accretion according to the equation 1 \& 2, if the jet power is not exceeding the $\dot{M}c^{2}$. Ghisellini et al. (2014) also found that the jet power in the Fermi gamma ray blazars is about 10 times the disk luminosity. We think it is possible that the jet efficiency is 10 times higher than the disk radiative efficiency in the equation 1, but it can not exceed the $\dot{M}c^{2}$ in the accretion dominated jet, otherwise the black hole spin could dominate the jet power: for instance, at the extreme case of $\Phi=\Phi_{MAD}$ in equation 3, we have:

\begin{equation}
P_{spin}\sim 2.5 (a)^{2} (\Phi/\Phi_{MAD})^{2} \dot{M}c^{2}=2.5(a)^{2}L_{disk}/\varepsilon
\end{equation}

So there is a possibility that the black hole spin-produced-jet may have `jet efficiency' $\eta=2.5a^{2}$ with compared to the equation 1, for extreme spinning black hole ($a=0.998$) $\eta$ can be $\sim 2.5$ (i.e. 250\% of $\dot{M}c^{2}$) about 10 times higher than the assumed maximum disk radiative efficiency of 30\% of $\dot{M}c^{2}$ (Ghisellini et al. 2014).

The radio-loud quasars are dominated by accretion disk with the linear correlation between jet power and disk luminosity (van Velzen \& Falcke 2013), which could be explained with the equation 1. For the `outliers' of BHXBs also shown a steeper correlation index between the jet and disk luminosity, a change at a critical accretion rate may be possible, at which the low
efficiency jet determined by the black hole spin would transit to the high efficiency jet dominated by inner disk accretion when the accretion rate
exceeds the critical rate (e.g. $\sim$ 0.01 Eddington luminosity).

In summary, the standard and `outliers' tracks of BHXBs could be related to a different coupling between jet power, mass accretion rate and black hole spin. For AGNs, there appear also subgroups divided by steep and flatter correlation indices. The correlation between jet power and disk luminosity cannot be simply unified with a single correlation slope from the BHXBs to AGNs, a third parameter (e.g. black hole spin) might be at work for the flatter correlation indices through a hybrid jet production from accretion and black hole spin. It is clear that the radio loud quasars which showing the nearly linear correlation between jet power and disk luminosity (e.g., van Velzen \& Falcke 2013) has to be dominated by the disk accretion as explained in the equation 1, and the contribution from black hole spin is small. We could still have a fundamental plane of black hole activity for the BHXBs and AGNs with the diverse (maybe two kinds of) correlation indices as discussed above, that the jets in the steep-index groups could be dominated by the accretion of black hole, while the jets in the flatter-index groups would be a hybrid production by the accretion and spin of black hole.

\section*{Acknowledgments}

Helpful comments from the reviewer are appreciated. This work is supported by the 973 Program 2015CB857100; the Key Laboratory of Radio Astronomy, Chinese Academy of Sciences; and the National Natural Science Foundation of China (No.11273050).

\end{document}